\documentclass[5p,times]{elsarticle}

\usepackage{hyperref,color} 

\journal{Mech. Res. Com.}

\newcommand{\glv}{\gamma_\mathrm{LV}}
\newcommand{\gsl}{\gamma_\mathrm{SL}}
\newcommand{\gsv}{\gamma_\mathrm{SV}}
\newcommand{\lin}{L_\mathrm{in}}
\newcommand{\lout}{L_\mathrm{out}}
\renewcommand{\d}{\mathrm{d}}
\newcommand{\V}{{\cal V}}
\renewcommand{\H}{{\cal H}}
\newcommand{\Heq}{\H^\text{eq}}
\renewcommand{\u}{\bm{u}}
\newcommand{\Tp}{T_\text{p}}

\graphicspath{{./Pics/}}
\usepackage{amsmath,bm}

\begin{document}

\begin{frontmatter}

\title{The heavy windlass: buckling and coiling of an elastic rod inside a liquid drop in the presence of gravity}

\address[idaCNRS]{Sorbonne Universit\'e, Centre National de la Recherche Scientifique, UMR 7190, Institut Jean Le Rond d'Alembert,  F-75005 Paris, France}
\address[idaSVI]{Surface du Verre et Interfaces, UMR 125 CNRS/Saint-Gobain, F-93303 Aubervilliers, France}

\author[idaCNRS]{Herv\'e Elettro} 
\author[idaCNRS,idaSVI]{Arnaud Antkowiak} 
\author[idaCNRS]{S\'ebastien Neukirch} \ead{sebastien.neukirch@upmc.fr}

\begin{abstract}
A liquid drop sitting on an elastic rod may act as a winch, or windlass, and pull the rod inside itself and coil it. This windlass effect has been shown to be generated by surface tension forces and to work best for small systems. Here we study the case where the drop is large enough so that its weight interferes with surface tension and modifies the windlass mechanics.
\end{abstract}


\end{frontmatter}


\section{Introduction}

Windlasses are winches used to pull weight or tighten sails on boats. They typically provide tension in a rope as it is wound around a cylinder, thereby transferring rotation momentum into translation momentum.
In 1989, Vollrath \& Edmonds \cite{Vollrath1989} proposed that the water droplets present along the threads in a spider web were acting as tiny windlasses, providing tension to the web, helping the structure to sustain loads ({\em e.g.} wind), and preserving its integrity.
Few years ago this windlass effect of small drops on micronic threads has been shown to be generated by capillary forces \cite{Elettro2016In-drop-capillary}: the affinity of the thread material for water is strong and surface tension acts against the elasticity of the thread, eventually buckling and coiling it.
Spider thread windlasses are yet another example of elastocapillarity, the study of the interplay between fluid forces and elasticity of solids \cite{Roman-Bico2010,Andreotti2016,Bico-2018}. Former examples include the bending of elastic plates around liquid drops \cite{Py-Reverdy2007},  the buckling of biofilaments inside liquid drops \cite{cohen+mahadevan:2003}, and the wetting of fiber arrays \cite{bico+al:2004,Duprat2012,Sauret2017} .
Previous works on the capillary windlass effect have focused on parameter values for which gravity could be discarded, namely for very small systems where the weight of the drop is much smaller than capillary forces, see {\em e.g.} \cite{Elettro2017,Schulman2017}.

Here we investigate the windlass system in the presence of gravity and show how its mechanics is changed when the weight of the drop is taken into account.
We introduce a simple analytical model and compute the bifurcation diagram of the system. We then perform experiments to test our theoretical predictions and show that gravity hinders the activation of the windlass mechanism. 

\section{Model} \label{section:model}
%
%
%
%
\begin{figure}[ht]
\begin{center}
\includegraphics[width=0.45\textwidth]{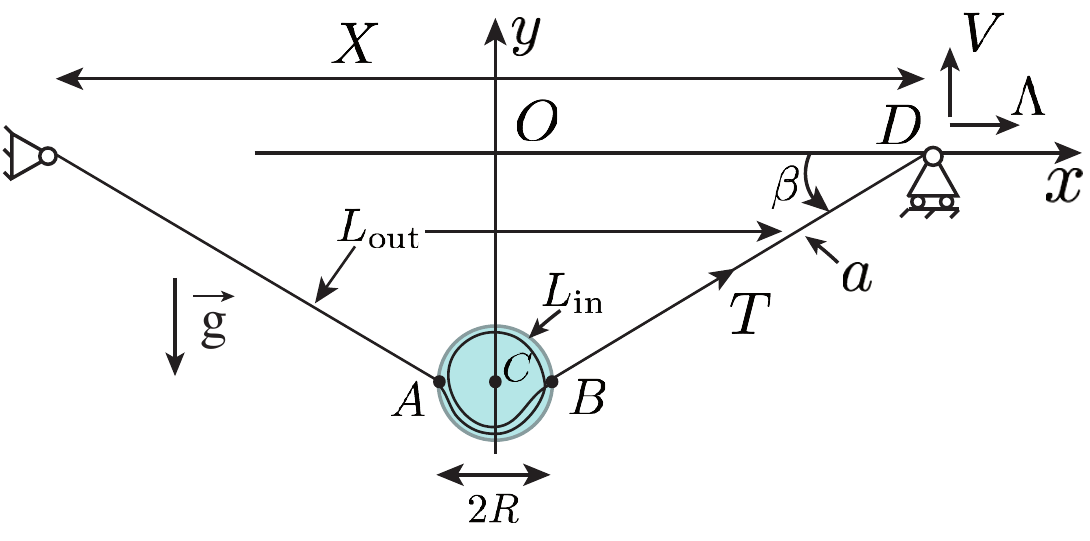}
\end{center}
\caption{A heavy drop of radius $R$ on a coilable elastic rod with circular cross-section of radius $a$ and total length $L=\lin+\lout$. The sagging angle $\beta$ is measured as the end-to-end distance $X$ is varied. The tension of the rod outside the drop is called $T$. Depending on the value of $X$, the rod inside the drop is either straight or coiled. The two points A and B where the rod enters/exits the drop are called meniscus points.}
\label{fig:model-notations}
\end{figure}
We consider an elastic rod of length $L$ and circular cross-section of radius $a$, made with an elastic material of Young's modulus $E$. The rod  then has a flexural rigidity $E \, I$, where $I=\pi a^4/4$ is the second moment of area of the section.
A liquid drop, of volume $4/3 \, \pi \, R^3$, is sitting astride the elastic rod. In the experimental setup we use systems with {\em e.g.} $a \sim 1 \mu$m, $R \sim 0.1$mm, $L \sim 1 $cm. In this case, the shape of the drop stays approximately spherical \cite{Carroll1976}.
The elastic rod is held at both extremities with pinned joints, see Figure \ref{fig:model-notations}, and we study the behaviour of the system as the right end is brought toward the left end, that is as the end-to-end distance $X$ is decreased.
In the present case where gravity is accounted for, the drop then goes down and the systems adopts a V shape, see Figure \ref{fig:model-notations}.
Nevertheless, as we deal with a sub-millimetric system, surface effects are coming into play and the affinity of the rod with the liquid has to be taken into account.
We do this by considering the energies of the three different interfaces present in the system. Per unit of area, we call 
$\gsv$ the energy of the solid-air interface, 
$\gsl$ the energy of the solid-liquid interface, and
$\glv$ the energy of the liquid-air interface.
In the case where the rod material has a stronger affinity with the liquid than with air, {\em i.e.} $\gsl < \gsv$, the rod may enter the liquid, that is, buckle and then coil inside the drop. Buckling will interfere with the simple V shape response mentioned earlier.
In order to calculate the behavior of the system, we write down its total potential energy and minimize it.
\begin{figure}[ht]
\begin{center}
\includegraphics[width=0.45\textwidth]{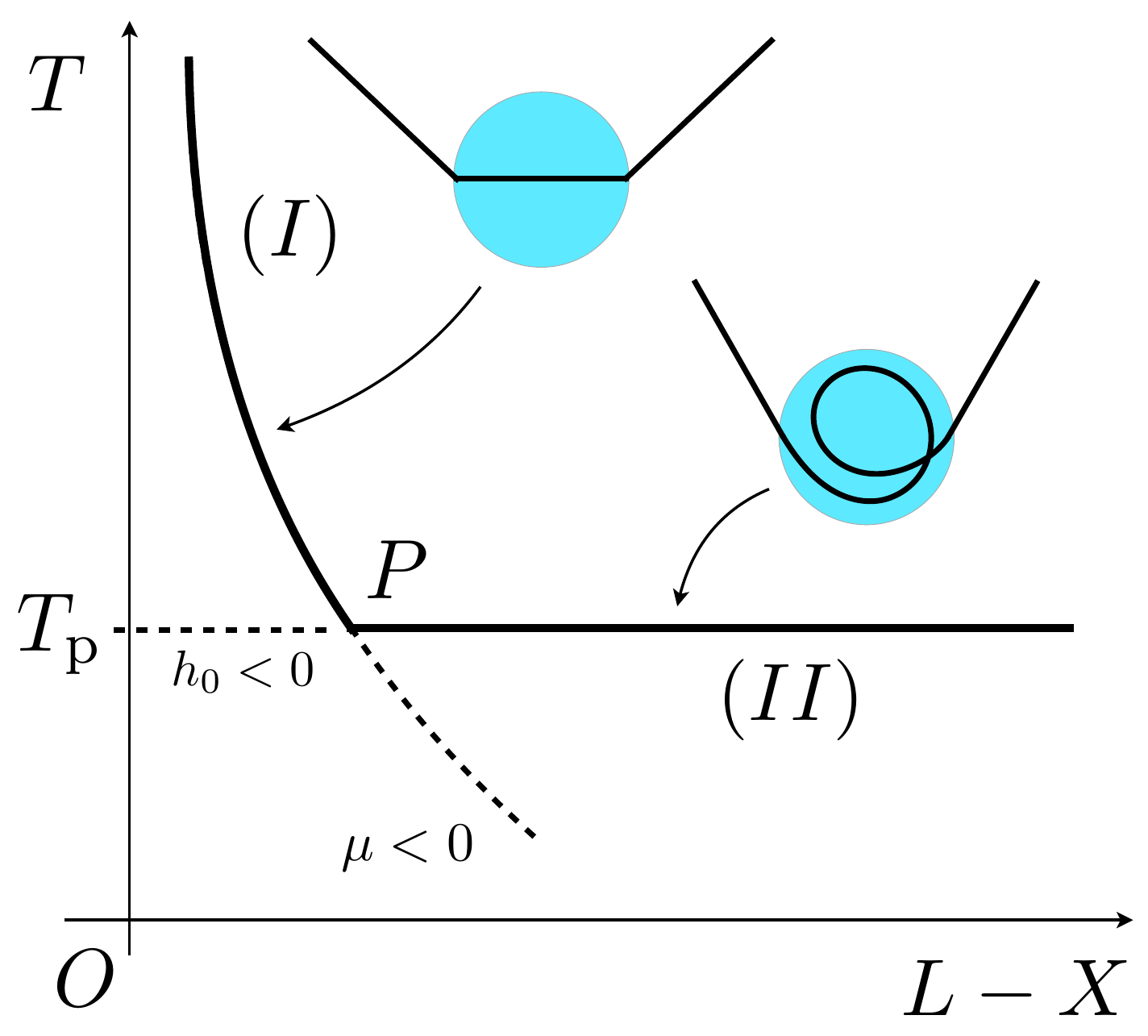}
\end{center}
\caption{Bifurcation diagram for the model presented in Section \ref{section:model}. The tension $T$ is plotted as a function of the end-shortening $L-X$. The diagram comprises two branches. Branch ($I$) corresponds to configurations where the system is unbuckled and behaves as a thread sagging under the weight of the drop, with the tension $T$ decreasing as the end-shortening is increased. Branch ($II$) corresponds to configurations where the elastic rod coils inside the drop and $T$ is constant, insensitive to the end-shortening.}
\label{fig:phase-diag}
\end{figure}

We first consider the potential energy of the weight of the drop $\V_g= M g Y_c$, where $Y_c$ is the center of the (spherical) drop, $M$ its mass, and $g$ the acceleration of gravity. We add the energy $\V_\gamma = 2 \pi a \lin \gsl + 2 \pi a \lout \gsv$ for the solid-liquid and solid-air interfaces, where $\lin$ is the total contour length of the rod inside the drop and $\lout$ is the total rod length outside the drop. Since in the present model we do not let the liquid drop deform, the energy for the liquid-air interface $4 \pi R^2 \glv$ is constant and therefore discarded. The bending energy of the elastic rod is $\frac{1}{2} \, E I \int_0^L \kappa^2(s) \, \d s$, where $\kappa(s)$ is the curvature of the rod and $s$ the arc-length along the rod. We perform the following simplifications: (i) the curvature outside the drop is considered zero, (ii) before buckling the curvature inside the drop is considered to vanish and the wetted length to be $\lin=2R$, while (iii) once coiling has started, $\lin > 2R$ and the coiled rod has curvature $1/R$. Under these assumptions the bending energy is written $\V_\kappa = \frac{1}{2} (\lin -2R) \, \frac{EI}{R^2}$.
Adding the different terms we end up with a total potential energy
\begin{align}
\V &(\lin,\lout,Y_c,\beta) = \V_g + \V_\gamma + \V_\kappa \nonumber \\
   & = MgY_c + 2\pi a \lin \gsl + 2\pi a \lout \gsv + (\lin -2R) \frac{EI}{2R^2}
\end{align}
We minimize this energy under the following constraints.  Extensional deformations are neglected so the total length $L$ is constant with $L=\lin+\lout$. Next, as we work with displacement controlled boundary conditions, we deal with fixed $X=2R + \lout \cos \beta $. Additionally the position of the center of the drop is given by $2Y_c+\lout \sin \beta=0$. Finally the wetted length $\lin$ cannot be smaller than $2R$, that is we have an inequality constraint $h_0 = \lin - 2R \geq 0$.
We use the first constraint to eliminate the variable $\lout$. Next, we cope with the second constraint  by introducing a
constraint function $h_1=(L-\lin) \cos \beta - X=0$ and a Lagrange multiplier  $\Lambda$, and with the third constraint by introducing a constraint function $h_2=2Y_c+[L-\lin] \sin \beta=0$ and a Lagrange multiplier $V$.
Finally the inequality constraint $h_0 \geq 0$ is treated by introducing a positive $\mu \geq 0$ multiplier and considering the Kuhn-Tucker condition $\mu \, h_0 = 0$.
Consequently we work with the Lagrangian ${\cal L}$ in the three-dimensional space $\u=(\lin,Y_c,\beta)$
\begin{align}
{\cal L}(\lin,Y_c,\beta)  &= \V -  \Lambda \left( [L-\lin] \cos \beta - X \right) \nonumber \\
             &~~~~~~ -  V \left(2Y_c+[L-\lin] \sin \beta \right)   - \mu (\lin -2R)
\end{align}
The necessary conditions for which the energy $\V$ is minimum are known as Kuhn-Tucker conditions \cite{Luenberger1973} and read
\begin{equation}
\frac{\partial \cal L}{\partial \u} = \bm{0} \, , \quad \mu \geq0 \, , \quad  \mu \, h_0 =0
\label{KT}
\end{equation}
%
We call these conditions equilibrium conditions.
Sufficient conditions to have a minimum, involving the Hessian matrix $\H$ with 
$\H_{ij}={\partial^2  {\cal L}}/({\partial u_i \, \partial u_j})$
\begin{equation}
\label{hessien}
\H = \left( 
\begin{array}{ccc}
0 & 0 & V \cos \beta - \Lambda \sin \beta \\
0 & 0 & 0 \\
V \cos \beta - \Lambda \sin \beta & 0 & (L-\lin) \, (\Lambda \cos \beta + V \sin \beta)
\end{array}
\right)
\end{equation}
will be called stability conditions.
Parameters are $R$, $EI$, $X$, $Mg$, $\Delta \gamma = \gsv - \gsl$ and
unknowns are $\lin$, $Y_c$, $\beta$, $V$, $\Lambda$, $\mu$.
\paragraph{Active constraint} ~\\
%
%
%
We first focus on solutions with $h_0=0$, that is before buckling happens. The solution to (\ref{KT}) is 
\begin{subequations}
\label{equa:equil-active-constr}
\begin{align}
\lin & = 2R \, , \:  Y_c=-\frac{X-2R}{2} \, \tan \beta \, , \:  \beta = \arccos \frac{X-2R}{L-2R} \\
V   & = Mg/2 \, , \: \Lambda= \frac{Mg}{2\tan \beta}
\end{align}
\end{subequations}
The Lagrange multipliers $\Lambda$ and $V$ are identified to be the horizontal and vertical reactions of the pinned joints.
Noting $T=\Lambda \cos \beta + V \sin \beta$ the tension in the rod outside the drop, we have 
\begin{equation}
T =\frac{Mg}{2 \sin \beta} ~ \text{ and } ~ \mu=T-\Tp
\end{equation}
%
where $\Tp=2\pi a \Delta \gamma -\frac{EI}{2R^2}$ and the condition $\mu \geq 0$ yields $T \geq \Tp$. The force $2\pi a \Delta \gamma$ corresponds to the compressive capillary force applied on the rod at the meniscus points $A$ and $B$, see Figure \ref{fig:model-notations}.

Starting from $X=L$ and decreasing $X$, the system follows Branch ($I$), drawn in Figure \ref{fig:phase-diag}. 
At first, when $X=L$, the tension $T$ outside the drop is infinite, as in any perfectly taut, horizontal string holding a weight. The tension then decreases down to the buckling point $P$, where $T=\Tp$.
The remaining, $T < \Tp$, of Branch $(I)$ has $\mu<0$ and therefore corresponds to configurations that do not fulfill equilibrium conditions (\ref{KT}).

Next, we test the stability of configurations in the upper part of Branch $(I)$. The Hessian matrix (\ref{hessien}) has to be evaluated 
for $\u = \u^{eq}$, where $\u^{eq}$ is given by (\ref{equa:equil-active-constr}). We find
%
%
\begin{equation}
\label{hessienEQ}
\Heq = \left( 
\begin{array}{ccc}
0 & 0 & 0 \\
0 & 0 & 0 \\
0 & 0 & (L-\lin) \, T
\end{array}
\right)
\end{equation}
A sufficient condition for stability is that 
$\delta \u \cdot \Heq \cdot \delta \u > 0$ for all $\delta \u=(\delta \lin, \delta Y_c, \delta \beta)$, where $\delta \u$
is a small variation about $\u^\text{eq}$ that has to be perpendicular to the three vectors $\partial h_1 / \partial \u$, $\partial h_2 / \partial \u$, and $\partial g / \partial \u$, see \cite{Luenberger1973} for more details. In the present case there is no variation perpendicular to all three vectors, hence no admissible variation\footnote{This special case comes from the fact that we have to fulfill three constraints in a three dimensional space.}. Equilibriums in the upper part of Branch $(I)$ are consequently all stable.

\begin{figure}[th]
\begin{center}
\includegraphics[width=0.95\columnwidth]{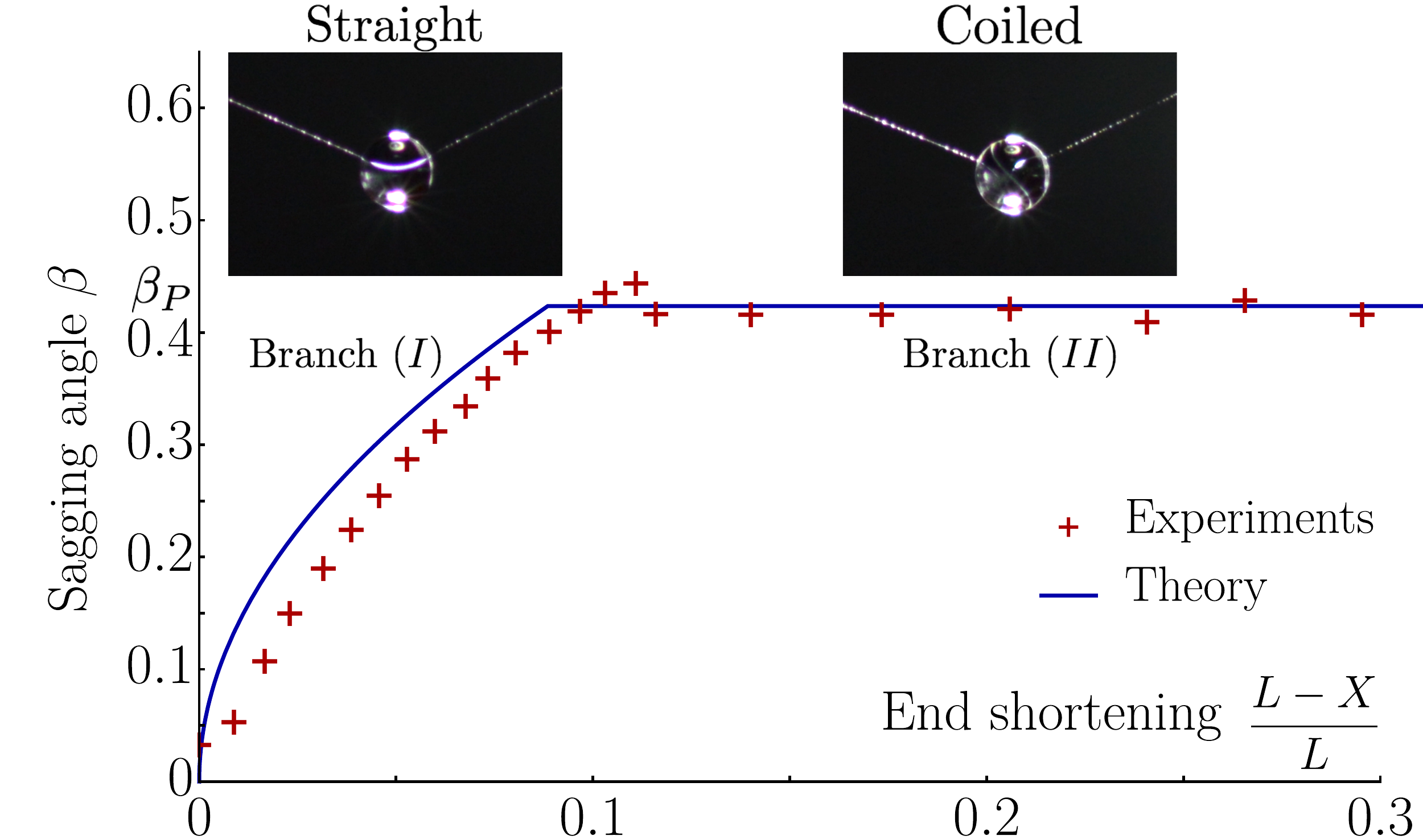}
\end{center}
\caption{Experimental bifurcation diagram for the windlass system with $Mg/(2\Tp)=0.46$. The sagging angle $\beta$ is recorded as the end-shortening $L-X$ is increased. The data clearly shows the two branches introduced in the theory and the flat, plateau, response in the second (coiling) regime, where the sagging angle is insensitive to the end-shortening. Theoretical predictions are also drawn and compare well with experimental points.}
\label{fig:beta-expe}
\end{figure}
\paragraph{Passive constraint} ~\\
%
%
%
We now describe solutions with $h_0>0$, that is $\lin>2R$. The solution to (\ref{KT}) is 
\begin{subequations}
\label{equa:equil-passive-constr}
\begin{align}
\lin &=L-\frac{X-2R}{\cos \beta}\, ,  \: Y_c=-\frac{X-2R}{2}  \, \tan \beta \, , \: T=\Tp \, , \:  \mu=0  \\
\beta & =\beta_\text{p}=\arcsin \frac{Mg}{2 \Tp} \, , \: V   =Mg/2 \, , \: \Lambda=\Tp \cos \beta  
\end{align}
\end{subequations}
The rod is steadily coiling inside the drop and the system evolves with constant $\beta$ and constant tension $T$.
In the decreasing $X$ experiment described earlier, as we reach the buckling point $P$ the system bifurcates on Branch $(II)$. Note that the part of the line $T=\Tp$ that lies before $P$ is such that $\lin < 2R$ and is therefore not physical.

Stability of solutions involve the same Hessian matrix (\ref{hessienEQ}) and we have to compute the sign of $\delta \u \cdot \Heq \cdot \delta \u$ for every $\delta \u$ perpendicular to both $\partial h_1 / \partial \u$, $\partial h_2 / \partial \u$ (the inequality constraint is now inactive). The subspace of admissible variation has dimension one and is given by 
$\delta \u = ([L-\lin] \sin \beta , [L-\lin]/2,-\cos \beta)$. We have $\delta \u \cdot \Heq \cdot \delta \u=\cos^2 \beta \, (L-\lin)\, T$, which is strictly positive as long as $T=\Tp >0$. Branch $(II)$ is then stable as long as the capillary force $2\pi a \Delta \gamma$ is larger than the coiling force $\frac{EI}{2R^2}$.

\section{Experiments} \label{section:experiments}
\begin{figure}[th]
\begin{center}
\includegraphics[width=0.95\columnwidth]{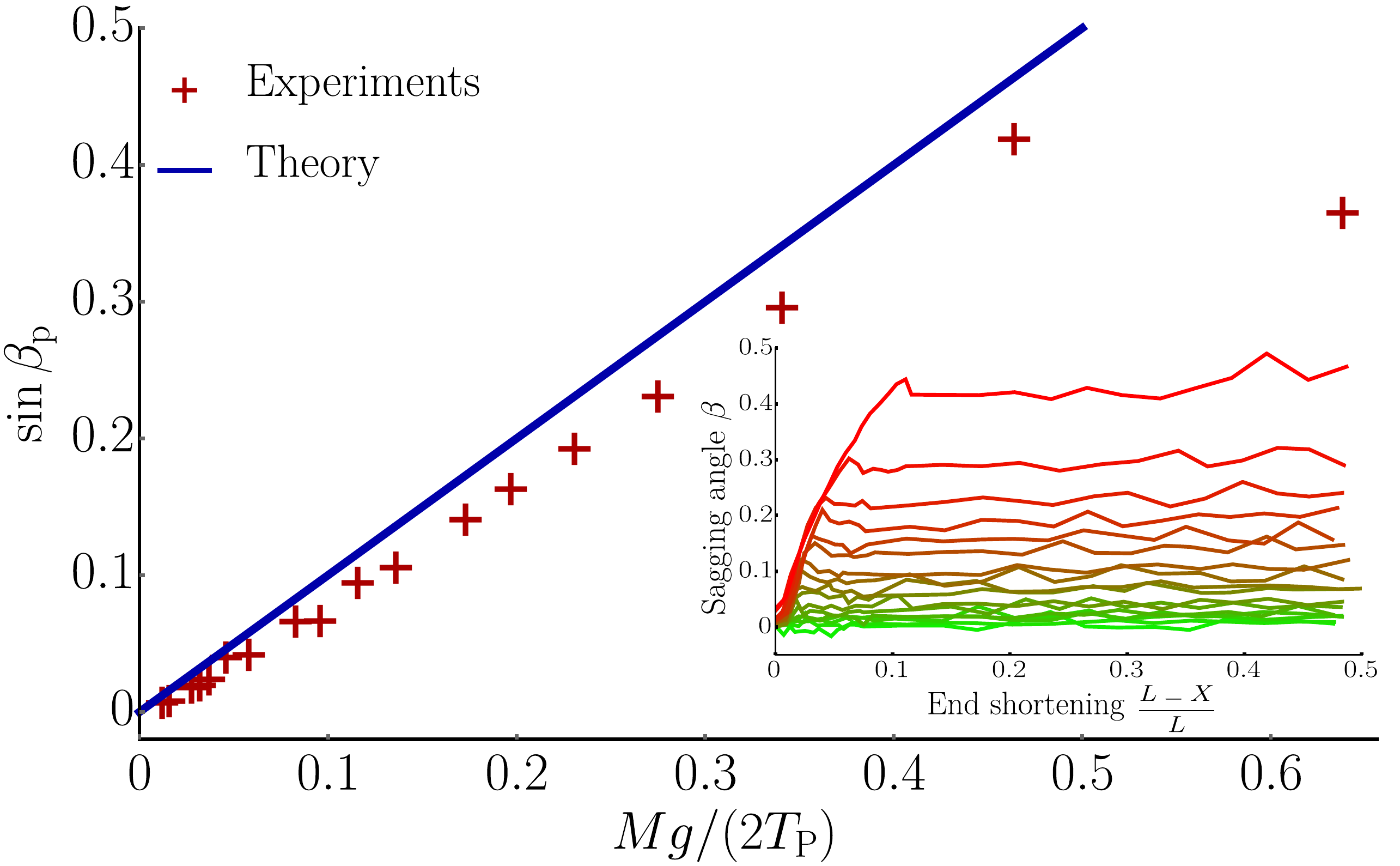}
\end{center}
\caption{Experimental data for the plateau value $\sin \beta_\text{p}$ as a function of the parameter $Mg/(2\Tp)$.
17 different diameters have been used $2R=$83.3, 92.0, 110.4, 115.5, 122.1, 131.1, 141.8, 159.8, 167.8, 178.5, \
188.2, 204.1, 213.2, 225.0, 238.5, 256.2, 284.0 $\mu$m. (Please note the presence of a 18th point, at $Mg/(2\Tp)=0.64$, corresponding to a rod with larger diameter, see Caption of Figure \ref{fig:hysteresis_gravity} for more informations).
(Inset) Experimental angle $\beta_\text{p}$ as function of end-shortening $L-X$ for the 17 values of the 
parameter $Mg/(2\Tp)$, ranging from 0.01 (green) to 0.46 (red).}
\label{fig:plateau-beta}
\end{figure}

%
%
%
Experiments were performed with a Thermoplastic Poly-Urethane (TPU) rod, of diameter $2a=2 \pm 0.7$  $\mu$m, produced by melt spinning, and Rhodorsil V1000 silicone oil droplets (density $\rho=960$ kg/m$^3$). Young's modulus for TPU was measured to be $17 \pm 2$ MPa, see \cite{Elettro2017} for more details.
For TPU and silicone oil, the surface tension $\Delta \gamma = \glv \cos \theta_Y$ was measured to have $\glv= 21.1$ mN/m and $\theta_Y = 23$ degrees. The value $g=9.81$ m/s$^2$ was used for the acceleration of gravity.
Care was taken to position the drop as close to the center of the rod as possible, as confirmed by the symmetric tilt of the two straight halves of the outside rod. 
The ratio of the weight $Mg$ to twice the plateau tension $2\Tp$ naturally arises from the model and we use it in the following.
%
%
%
Liquid drops with diameter $2R$ ranging from 83.3 to 284 $\mu$m were used, yielding a ratio $Mg/(2 \Tp)$ ranging from $0.01$ to $0.46$.
The end-to-end distance $X$ was decreased at a speed of $12$ $\mu$m/s and no dynamic effect was observed. We followed the position of the drop using particle tracking at the rate of one image every second, and we recorded the angle $\beta$ as a function of the end-shortening $L-X$, see Figure~\ref{fig:beta-expe}.
We also recorded the value of the constant angle $\beta_P$ in the coiling regime for each different drop radius, and plotted $\sin \beta_P$ as a function of $Mg/(2 \Tp)$, see figure (\ref{fig:plateau-beta}).

\section{Discussion and Conclusion} \label{section:conclusion}
%
%
%


\begin{figure}[ht]
\begin{center}
\includegraphics[width=0.45\textwidth]{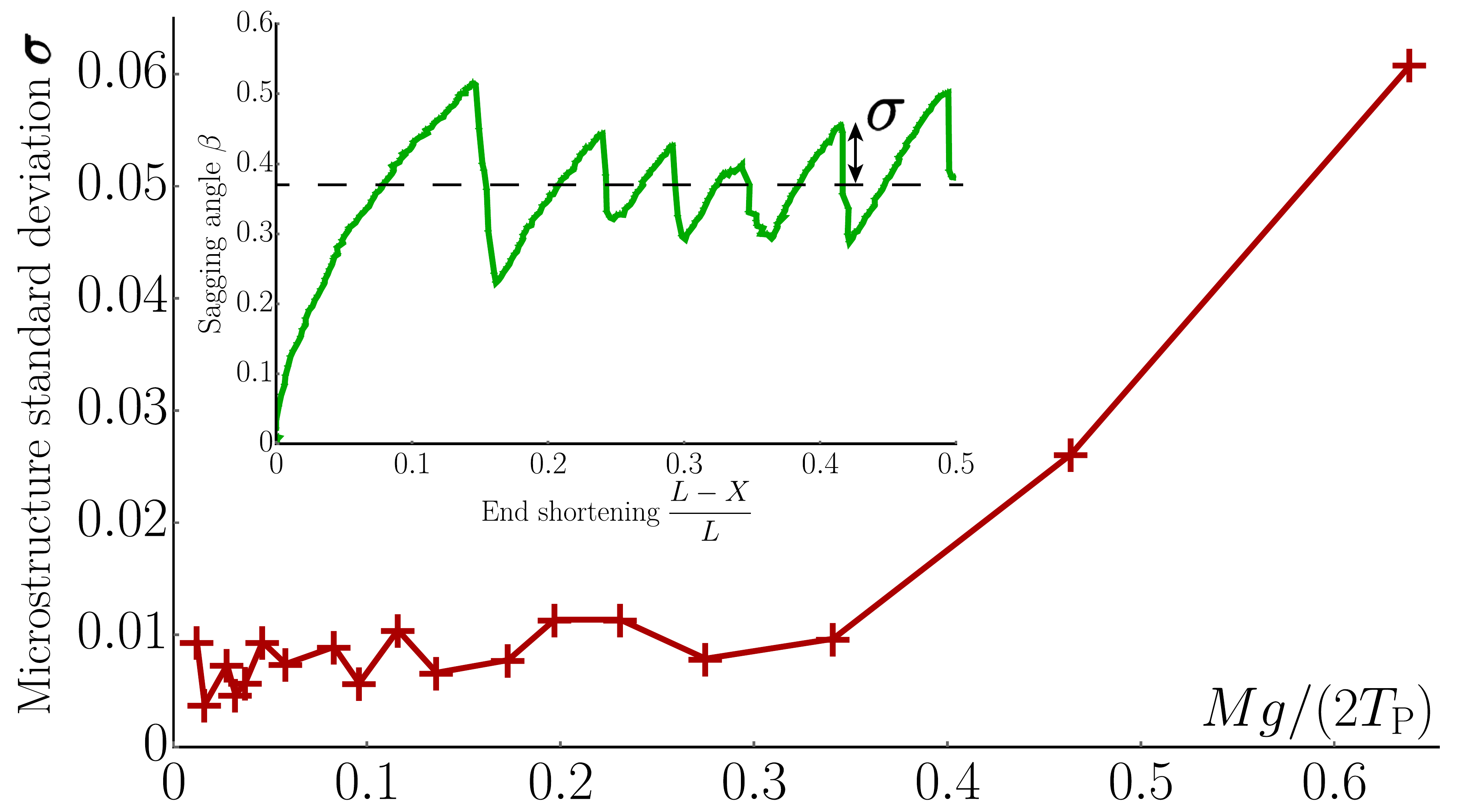}
\end{center}
\caption{The `thickness' $\sigma$ of Branch $(II)$ as function of $Mg/(2\Tp)$. Inset: Details of the experimental bifurcation curve for $Mg/(2\Tp)=0.64$ ($2a=8.3 \pm 0.7$ $\mu$m and $2R=497 \pm 2$ $\mu$m), showing the microstructure of Branch ($II$).}
\label{fig:hysteresis_gravity}
\end{figure}

Experiments clearly show a two-regime response of the system, corresponding to the two bifurcation branches introduced in Section \ref{section:model}. In Figure \ref{fig:beta-expe} we see that the angle $\beta$ starts by increasing as $X$ is decreased, as predicted by Eq.~(\ref{equa:equil-active-constr}). In this first regime, the system behaves as a flexible thread sagging under the application of a dead weight, with the sag increasing as the tension in the thread decreases. As the tension reaches the threshold $\Tp$ the rod buckles in the drop which no longer acts as a dead weight but rather as an active winch, reeling in and coiling more and more rod length as the end-to-end distance $X$ is decreased.
This second regime is characterized by a flat, plateau, response where the angle $\beta$ no longer depends on $X$. In addition to the experimental data, we draw in Figure \ref{fig:beta-expe} the two branches predicted by the theory and find a good agreement between experiments and theory. We also note on Figure~\ref{fig:beta-expe} that, experimentally, the highest point of Branch $(I)$ is higher than the plateau value $\beta_P$. This mismatch is due to the subcritical nature of the buckling transition, which is studied in \cite{Elettro2017} in the absence of gravity and remains in the presence of gravity.
Consequently the buckling transition is here associated with a small jump, a sudden decrease of the angle $\beta$ and a sudden increase of the tension $T$.
Varying the drop radius, we experimentally record the plateau value $\beta_P$ and test Equation (\ref{equa:equil-passive-constr}) according to which $\sin \beta_P$ should be equal to $Mg/(2\Tp)$. We see in Figure~\ref{fig:plateau-beta} that the experimental data agrees with the linear dependance $\sin \beta_P = Mg/(2\Tp)$ but that a deviation starts to build for large values of $Mg/(2 \Tp)$.
This discrepancy is thought to be due to the spherical drop assumption in our model: experimentally, in order to equilibrate the hydrostatic pressure within the drop, gravity modifies the shape of the drop as well as  the location of the meniscus points $A$ and $B$ \cite{Lorenceau2004Capturing-drops}.

Finally, we note that experimentally Branch $(II)$ is not a mere straight line but carries a microstructure, shown in the inset of Figure~\ref{fig:hysteresis_gravity}. At several locations, the system jumps between two configurations with different values of the angle $\beta$  and the tension $T$. 
Upon coiling (decreasing $X$), during a jump,  $\beta$ decreases and $T$ increases, while upon uncoiling (increasing $X$) the situation would be reversed. 
%
We see in Figure \ref{fig:hysteresis_gravity} that the `thickness' $\sigma$ of Branch $(II)$ increases with the size of the drop and becomes large when $Mg/(2 \Tp)>0.5$. In this case the drop becomes so large that the geometrical hypotheses in our model (spherical assumption, location of meniscus points $A$ and $B$) break down. Again, this is illustrated by the $\beta_\text{p}$ value for the last point $Mg/(2 \Tp)=0.64$ which does not fit the linear regime, see Figure \ref{fig:plateau-beta}.


In the case the weight of the drop vanishes, $M \,g \to 0$, the bifurcation diagram of Figure \ref{fig:phase-diag} is modified and the buckling point $P$ approaches the vertical axis ($X_\text{p} \to L$, $\beta_\text{p}\to 0$), keeping the same $\Tp$ value. 
As we are not considering the intrinsic extensibility of the rod, in this limit $M \,g \to 0$, Branch $(I)$ only comprises the upper part $T>\Tp$ of the vertical axis, and buckling takes place as soon as some non-zero end-shortening $L-X$ is introduced. The weight of the drop therefore induces a delay on windlass activation: one has to reach a finite end-shortening $L-X_\text{p}$ to induce activation. For small weights, the end-shortening at activation grows as 
\begin{equation}
\frac{L-X_\text{p}}{L} \simeq \frac{1}{2} \, \left(\frac{Mg}{2\Tp} \right)^2
\end{equation}
For large weights, activation is prevented as soon as $\beta_\text{p}$ reaches $\pi/2$, that is as $Mg$ reaches $2\Tp$. 
Finally we note that once the windlass is activated, the tension in the rod outside the drop is independent of the weight $Mg$: we have $T=\Tp$ even if, {\em e.g.}, $Mg=0$.

\section*{Acknowledgements}
%
%
%
%
%
We gratefully thanks Natacha Krins, David Grosso, C\'edric Boissi\`ere, Sinan Haliyo, and Camille Dianoux for help with experiments and for discussions.
The present work was supported by the french Agence Nationale de la Recherche, grants ANR-09-JCJC-0022-01 and ANR-14-CE07-0023-01, by the city of Paris, grant `La Ville de Paris, Programme Emergence', and by the CNRS through a PEPS PTI program.

This communication is dedicated to the memory of G\'erard Maugin. He has provided me (SN) with continuous support and advices on my research activities. As an Emeritus member of our department, our offices were facing and I have seen him coming in everyday, sitting at his desk, mumbling in his beard, and working on new book projects until the end.

\bibliographystyle{elsarticle-num}
\bibliography{w5-gravite}
\end{document}